%% file: manuscript.tex
\documentclass[modern]{aastex701}

\usepackage{amsmath}
\usepackage{wasysym}
\input{shortcuts.tex}

\begin{document}


\title{Solar Hegemony:\\M-Dwarfs Are Unlikely to Host Observers Such as Ourselves}

\author[orcid=0000-0002-4365-7366,sname='Kipping']{David Kipping}
\affiliation{Columbia University, 550 W 120th Street, New York NY 10027}
\email[show]{dkipping@astro.columbia.edu}  

\begin{abstract}
\input{abstract.tex}
\end{abstract}

\keywords{exoplanets --- astrobiology}


\section{Introduction}
\label{sec:intro}

\input{intro.tex}

\section{Bayesian Introspection}
\label{sec:model}

\input{model.tex}

\section{Results}
\label{sec:results}

\input{results.tex}

\section{Discussion}
\label{sec:discussion}

\input{discussion.tex}



\section*{Acknowledgements}

Thank-you to Avi Loeb for a helpful conversation about his previous work on this topic, and to Jason Wright for constuctive comments on an early draft.
Special thanks to donors to the Cool Worlds Lab, without whom this kind of research would not be possible:
Douglas Daughaday,
Elena West,
Tristan Zajonc,
Alex de Vaal,
Mark Elliott,
Stephen Lee,
Zachary Danielson,
Chad Souter,
Marcus Gillette,
Jason Rockett,
Tom Donkin,
Andrew Schoen,
Mike Hedlund,
Ryan Provost,
Nicholas De Haan,
Emerson Garland,
The Queen Road Foundation Inc,
Ieuan Williams,
Axel Nimmerjahn,
Brian Cartmell,
Guillaume Le Saint,
Robin Raszka,
Bas van Gaalen,
Josh Alley,
Drew Aron,
Warren Smith,
Beauregard Burgess \&
Brad Bueche.

\bibliography{manuscript}{}
\bibliographystyle{aasjournalv7_surnames_only}



\end{document}

%% file: shortcuts.tex
\newcommand{\pdf}{\mathrm{Pr}}
\newcommand{\Mcrit}{M_{\mathrm{crit}}}
\newcommand{\Twin}{T_{\mathrm{win}}}
\newcommand{\tbirth}{t_{\mathrm{birth},i}}
\newcommand{\zams}{t_{\mathrm{ZAMS},i}}
\newcommand{\tams}{t_{\mathrm{TAMS},i}}
\newcommand{\tmax}{t_{\mathrm{max}}}

%% file: abstract.tex
With no firm evidence for life beyond our solar system, inferences about the population observers such as ourselves rests upon the Earth as a single input, at least for now. Whilst the narrative of our home as a ``humdrum'' system has become ingrained in the public psyche via Sagan, there are at least two striking facts about our existence which we know are certainly unusual. First, the stelliferous period spans ${\sim}10$\,Tyr - yet here we are living in the first 0.1\% of that volume. Second, over three-quarters of all stars are low-mass M-dwarfs, stars with no shortage of rocky habitable-zone planets - and yet, again, our existence defies this trend, previously dubbed the Red Sky Paradox. Two plausible resolutions are that a) stars below a certain mass, $\Mcrit$, do not produce observers, and, b) planets have a truncated temporal window for observers, $\Twin$, negating the longevity advantage of M-dwarfs. We develop a Bayesian model that encompasses both datums and jointly explores the two resolutions covariantly. Our analysis reveals that
1) the hypothesis that these observations are mere luck is disfavored with an overwhelming Bayes factor of ${\simeq}1600$;
2) some truncation of low-mass stars is indispensable, lowering $\Twin$ alone cannot well-explain the observations; and,
3) the most conservative limit on $\Mcrit$ occurs when fixing $\Twin=10$\,Gyr, yielding $\Mcrit>0.34$\,$M_{\odot}$ [$0.74$\,$M_{\odot}$] to 2\,$\sigma$ [1\,$\sigma$]. Our work challenges the tacit assumption of M-dwarfs being viable seats for observers and, indirectly, even life.

%% file: intro.tex
Humanity's efforts to search for signs of technology beyond our solar system have, thus far, revealed no compelling evidence. In truth, these efforts have been extremely limited in scope, curtailed by minimal funding, and are estimated to equate to sampling a large bathtub of water in comparison to the Earth's oceans \citep{tartar:2010,wright:2018}. Any inferences about the existence of observers such as ourselves is thus limited to the sole example we know of - us. Many might despair at such a state of affairs and argue that a single example is devoid of any useful information \citep{monod:1971}. But in truth, zero information is defined by zero data points (not one). Indeed, inferences from a single datum have less formally guided astronomers for generations already, such as predicting that other stars were Suns \citep{galileo:1610}, the plurality of galaxies \citep{kant:1755} and that other stars would have planets \citep{sagan:1966}.

Accordingly, there have numerous efforts to apply such thinking to the question of life. The probability and timescale for abiogenesis have been subject to particular attention in this regard, kickstarted by the seminal work of \citet{lineweaver:2002}. \citet{spiegel:2012} explored the subtle influence of choice of priors has on this problem and \citet{kipping:2020} later folded in the covariance with evolutionary timescale.

As a general principle of statistics, a random sample drawn from some unknown distribution is unlikely to be a tail-end outlier. This reasoning informs the Copernican Principle, which argues that amongst all observers, our experience is not privileged and should thus be a representative sample; an idea which has become a cornerstone of modern cosmology \citep{clarkson:2008,zhang:2011}. Accordingly, the Earth should be a representative example of what a planet inhabited by observers looks like. It is tempting to go through every peculiarity of the Earth and argue that this is how all inhabited planets must therefore look, the so-called ``Rare Earth Hypothesis'' \citep{ward:2000}. But in reality, we have no idea how rare or common each of these features are, for example the prevalence of large moon around rocky planets remains beyond observational capabilities \citep{kipping:2022}.

Nevertheless, there are two features of our home, both in space and time, that are incontrovertibly unusual. The first is that approximately three-quarters of all stars are M-dwarfs \citep{reyle:2021} and yet we do not live around one. This puzzle is compounded by recent exoplanet surveys which reveal an abundance of Earth-sized worlds in the habitable-zones of M-dwarf stars \citep{dressing:2015}. \citet{haqq:2018} first highlighted this quandary and later \citet{redsky} remedied a normalization problem of that earlier approach, highlighting possible resolutions to this ``Red Sky Paradox'' (RSP). However, a limitation of \citet{redsky} is that the star formation history was not accounted for and stellar ages are crudely averaged over spectral types.

The second puzzle is that we live so early in the stelliferous period of our Universe, an epoch which should span $\mathcal{O}[10^{13}]$ years \citep{adams:1997}. As highlighted by \citet{redsky}, this is of course intimately related to RSP since M-dwarfs are the protagonists for most of this period. \citet{loeb:2016} considered this puzzle using an astrophysical prescription for star formation history, the initial mass function and main sequence lifetimes. Although not a Bayesian analysis, \citet{loeb:2016} suggested the resolution that habitability around M-dwarfs might be suppressed.

The two puzzles are clearly two sides of the same coin and thus in this work we attempt to fuse the more complete astrophysical model of \citet{loeb:2016} (with some improvements) and the rigorous Bayesian philosophy of \citet{redsky} to tackle this problem, which would of course also dissolve RSP.

\citet{redsky} formally lists four resolutions to RSP, but resolution I is simply luck and thus is intrinsically built into a probabilistic model such as that defined in this work. Further, resolutions II and IV both equate to inhibiting the development of observers around M-dwarfs (II is biological and IV is lack of Earths), and thus we combine them in this work. No specific mechanism is assumed here, but plausibly this could be due to their distinct radiation environment \citep{rugheimer:2015} or atmospheric erosion \citep{segura:2010}, for example. This leaves us with one remaining solution (resolution III of \citealt{redsky}) - that M-dwarfs planets have truncated windows for observers as compared to their host star's lifetime, plausibly due to the cessation of active geology \citep{oneill:2016,cheng:2018}.

In this work, we conduct a Bayesian analysis of both possible explanations conditioned upon both datums. Our goal is to investigate their explanatory power and constrain the relevant parameters of each. We first introduce our Bayesian and astrophysical model in Section~\ref{sec:model}, before exploring the posteriors and comparing models in Section~\ref{sec:results}. Section~\ref{sec:discussion} considers the ramifications of our results, particularly in our search for life elsewhere. 

%% file: model.tex
\subsection{Formulating the Problem}

There are two puzzles about our existence, or really two data points we can feed into our Bayesian inference model. The first is the date of our existence, which we write as $t=t_0$ where $t_0$ is the current age of the Universe (we adopt $t_0=13.789$\,Gyr; \citealt{louis:2025}). The second datum is that our host star is a G2-type star, which we write as $M_{\star}=M_{\odot}$. Thus, we could write that our data, upon which our subsequent inferences will be explicitly conditioned on, is given by $\mathcal{D} = \{t=t_0,M_{\star}=M_{\odot}\}$.

Next, we require a model to explain the data with one or more parameters which we will attempt to infer, conditioned upon $\mathcal{D}$. As noted in Section~\ref{sec:intro}, this work focusses on two possible explanations. The first is that stars below some critical mass, here dubbed as $\Mcrit$, do not produce observers (although they may still produce simple life). We refer to this as the ``desolate M-dwarf hypothesis'' throughout. One might argue that a sharp cut-off like this is unrealistic, and perhaps prefer a two-parameter logistic function with a smooth activation function. Whilst certainly a reasonable stance, the extremely data-starved regime in which we operate here (just two data points) compels us to try to simplest conceivable model in order to make progress.

The second explanation we consider is that the temporal window for observers is not solely governed by the lifetime of its parent star, but that other processes intrinsic to the planet may cut it short (most plausibly geophysics). We refer to this as the ``truncated window hypothesis'' throughout. For the sake of simplicity, we again apply a broad brush model here that Earth-like planets have an intrinsic lifetime for observers of $\Twin$. Our task is thus to infer $\pdf(\Mcrit,\Twin|t=t_0,M_{\star}=M_{\odot})$ - the joint posterior distribution of $\Mcrit$ and $\Twin$.

\subsection{Bayesian Framing}

We can use Bayes' theorem to expand the joint posterior as

\begin{align}
\underbrace{\pdf(\Mcrit,\Twin|t=t_0,M_{\star}=M_{\odot})}_{\mathrm{joint\,\,posterior}} &= \frac{ \overbrace{\pdf(t=t_0,M_{\star}=M_{\odot}|\Mcrit,\Twin)}^{\mathrm{joint\,\,likelihood}} \overbrace{\pdf(\Mcrit,\Twin)}^{\mathrm{joint\,\,prior}} }{ \underbrace{\pdf(t=t_0,M_{\star}=M_{\odot})}_{\mathrm{evidence}} },
\label{eqn:bayes}
\end{align}

where we have annotated the various components in the standard parlance of Bayesian inference. To make progress, we assume that the prior probability distribution of $\Mcrit$ and $\Twin$ are independent such that $\pdf(\Mcrit,\Twin) = \pdf(\Mcrit)\pdf(\Twin)$. We cannot reasonably make the same assumption for the joint likelihood, but we can still make progress by expanding hierarchically as

\begin{align}
\pdf(\Mcrit,\Twin|t=t_0,M_{\star}=M_{\odot}) \propto&
\overbrace{ \pdf(M_{\star}=M_{\odot}|t=t_0,\Mcrit,\Twin)}^{ \mathcal{L}_m } \nonumber\\
\qquad& \underbrace{ \pdf(t=t_0|\Mcrit,\Twin) }_{ \mathcal{L}_t } \pdf(\Mcrit) \pdf(\Twin),
\label{eqn:posterior}
\end{align}

where we have labeled two likelihood functions, $\mathcal{L}_m$ and $\mathcal{L}_t$, that we need to calculate in what follows. Also, note how we have dropped the evidence term for the moment, since we can always compute this by integrating the joint posterior to ensure correct normalization.

\subsection{Priors}

We adopt a non-informative uniform prior for $\Mcrit$ i.e. all values are equally likely \textit{a-priori}, such that

\begin{equation}
\pdf(\Mcrit) =
\begin{cases}
\frac{1}{M_{\odot} - M_{H}} & \text{if } M_H \leq \Mcrit < M_{\odot},\\
0 & \text{otherwise},
\end{cases}
\label{prior:Mcrit}
\end{equation}

where $M_H$ is the hydrogen burning limit, the lowest mass star - here assumed to be $M_H = 0.08$\,$M_{\odot}$. Note how our prior stops at $M_{\odot}$ - we know for sure that Sun-mass stars can have observers around them so the prior cannot encompass $M_{\odot}$.

Similarly, we adopt a non-informative prior for $\Twin$, except this term spans several orders of magnitude (Gyr to Tyr) and thus a log-uniform prior is more appropriate.

\begin{equation}
\pdf(\Twin) =
\begin{cases}
\frac{\Twin^{-1}}{ \log(T_{\mathrm{win},\mathrm{max}}) - \log(T_{\oplus}) } & \text{if } T_{\oplus} \leq \Twin < T_{\mathrm{win},\mathrm{max}},\\
0 & \text{otherwise},
\end{cases}
\label{prior:Twin}
\end{equation}

where $T_{\oplus}$ is the current age of the Earth (a formal minimum for $\Twin$) and $T_{\mathrm{win},\mathrm{max}}$ is the maximum possible window considered. We pause to more strictly define $\Twin$. It represents essentially a death-date in this work, with respect to time when the host star was born. After this date, the planet cannot naturally develop observers.

\subsection{Likelihood $\mathcal{L}_t$}
\label{sub:liket0}

To evaluate the likelihood of an observer emerging at the present age of the Universe, $t_0$, we need to evaluate the number of viable seats for observers as a function of time. Recall that in our simplified model, the only parameter sculpting the viability of a star to host observers is $\Mcrit$. Accordingly, stars with $M_{\star}>\Mcrit$ are all equally likely to host observers. Once again, we acknowledge that more sophisticated models could be proposed, but the dearth of constraining data does not warrant such complications at this time.

We thus decided to create a numerical simulation of the chronology of the Universe during its stelliferous era. For $10^6$ stars, we first assign each a random birth date, $\tbirth$, by drawing this value from a probability distribution representing the star formation rate (SFR) of the Universe over time. For each star, we then assign a random mass, $m_i$, drawn from the initial mass function (IMF). We then step through the timeline in $\Delta t = 0.1$\,Gyr steps counting up how many of our stars satisfy our criteria for hosting observers. Specifically, we require that the $i^\mathrm{th}$ star can only host observers if, at time $t_j$, we have

\begin{enumerate}
\item $t_j > \tbirth + \zams + T_{\mathrm{delay}}$,
\item $t_j < \mathrm{min}[\tbirth + \tams, \tbirth + \Twin ]$,
\item $m_i > \Mcrit$,
\end{enumerate}

where $\zams$ ($\tams$) is the zero (terminal) age main sequence age of the $i^\mathrm{th}$ star, and $T_{\mathrm{delay}}$ is a universal, fixed parameter defining some delay time for observers to develop after their star is born. We adopt 3\,Gyr for this parameter - the time it took for the first multicellular life to emerge on Earth. Due to our TAMS lifetime formula discussed shortly, this effectively means that stars $M_{\star}>M_{\odot}$ will not live long enough to ever produce observers. Our simulation terminates after three times steps in a row yield no inhabited stars at all.

Our astrophysical model takes inspiration from \citet{loeb:2016}. In particular, we adopt the identical empirical model of the star formation rate per comoving volume as a function of redshift, $z$ - that of \citet{madau:2014}:

\begin{align}
\dot{\rho_{\star}}(z) &\propto \frac{(1+z)^{2.7}}{ 1 + ((1+z)/2.9)^{5.6} }.
\label{eqn:SFR}
\end{align}

Unlike \citet{loeb:2016}, we treat this is an unnormalized probability density function, which we first normalized, then computed the cumulative density function of and then solved the inverse density function using Newton's method. This allows us to draw a random (but representative) sample for the birth redshift of a star. Redshifts are then converted to time.

We also adopted the same TAMS function as \citet{loeb:2016}, which follows the \citet{adams:2004} tracks for low mass stars. A downside of these tracks is that they assume Solar metallicity, but despite this, they remain the most detailed calculations of the final phases of low mass stars. Accordingly, we have

\begin{equation}
\tams =
\begin{cases}
(10\,\mathrm{Gyr}) (m/M_{\odot})^{-2.5} & \text{if } 0.75\,M_{\odot} < m_i < 3\,M_{\odot},\\
(7.6\,\mathrm{Gyr}) (m/M_{\odot})^{-3.5} & \text{if } 0.25\,M_{\odot} < m_i \leq 0.75\,M_{\odot},\\
(53\,\mathrm{Gyr}) (m/M_{\odot})^{-2.1} & \text{if } 0.08\,M_{\odot} \leq m_i < 0.25\,M_{\odot}.
\end{cases}
\label{eqn:lifetimes}
\end{equation}

An important difference between our model and that of \citet{loeb:2016} is the inclusion of ZAMS, which we evaluated by interpolating the MESA Isochrones and Stellar Tracks (MIST) grids \citep{mist1,mist2} for Solar metallicity (for consistency with the TAMS calculation). Of course, another difference is the inclusion of the evolutionary delay time, $T_{\mathrm{delay}}$.

We also do not use the \citet{chabrier:2003} IMF but instead use the ``canonical'' IMF of \citet{kroupa:2024}. However, like \citet{loeb:2016}, we assume this is static with respect to redshift, largely due to the lack of a good alternative at the time of writing \citep{kroupa:2024}. The IMF is truncated to $\Mcrit$ in a given simulation, since there's little point simulating stars that can never produce observers in this analysis.

\begin{figure}
\begin{center}
\includegraphics[width=15.5 cm]{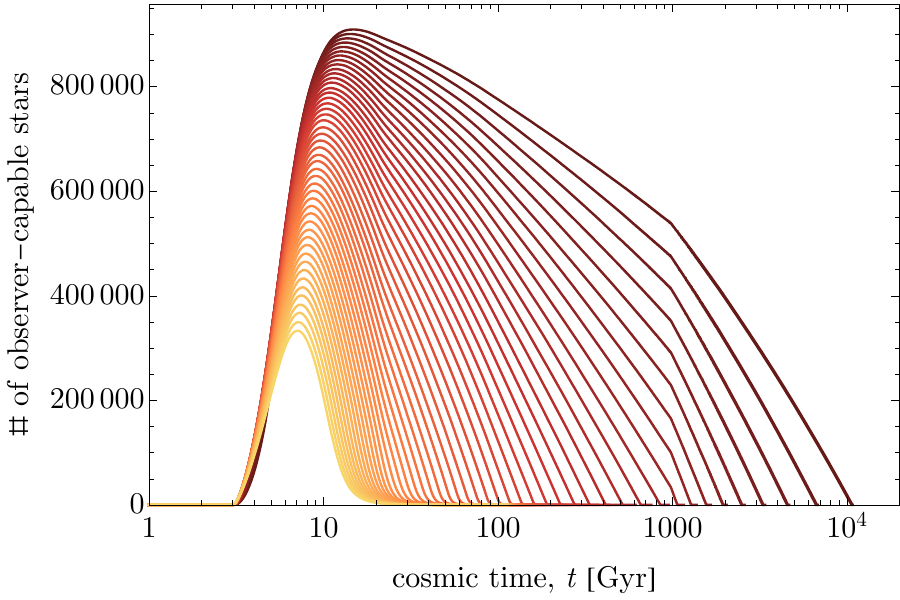}
\caption{\emph{
Number of stars capable of hosting observers (amongst a sample of $10^6$) as a function of time. The darkest track assumes that all stars are equally capable of hosting life during their MS lifetime. Each lighter track sequentially truncates the lowest mass star capable of hosting observers in 0.02\,$M_{\odot}$ steps. One can see that removing low-mass stars gradually relaxes the tension as to why we live so early in the stelliferous period.
}} 
\label{fig:masstracks}
\end{center}
\end{figure}

Figure~\ref{fig:masstracks} shows an example of the number of stars capable of hosting observers as a function of time, computed using the prescription outlined above and for a selection of $\Mcrit$ values (and assuming $\Twin = T_{\mathrm{win},\mathrm{max}}$). We find that even with $\Mcrit=0.08$\,$M_{\odot}$, zero stars (out of $10^6$) are left by 10.7\,Tyr and thus we set $T_{\mathrm{win},\mathrm{max}} = 10$\,Tyr, since beyond this point it would have negligible impact. The curves in Figure~\ref{fig:masstracks} indeed directly yield $\mathcal{L}_t = \pdf(t=t_0|\Mcrit,\Twin)$ by calculating said curve for a given choice of $\Mcrit$ and $\Twin$, interpolating the curve, normalizing into a proper distribution and then evaluating at the point $t=t_0$.

\subsection{Likelihood $\mathcal{L}_m$}
\label{sub:likeMsun}

We next turn to evaluating $\mathcal{L}_m = \pdf(M_{\star}=M_{\odot}|t=t_0,\Mcrit,\Twin)$. Due to the triple conditional, the task is somewhat simplified versus the $\mathcal{L}_t$. Specifically, rather than worrying about the distribution of masses at each of our time steps, we only care about the mass distribution at $t=t_0$. When our time step hits 13.8\,Gyr, we thus instruct our simulation to export the masses of all stars which satisfy our criteria for an observer. This will of course be bespoke to a given choice of $\Mcrit$ and $\Twin$, as before with $\mathcal{L}_t$. The list of masses is then converted into a distribution using kernel density estimation (KDE) with a Gaussian kernel, and finally we evaluate the resulting probability density function at $M_{\odot}$ yielding $\mathcal{L}_m$.

\subsection{Calculated Grids}

Our two likelihood functions are sensitive to two inputs, $\Mcrit$ and $\Twin$. We thus proceed by calculating these likelihoods across a grid in $\{\Mcrit,\Twin\}$ spanning the prior volume, from which we can then build a smooth interpolated function. As eluded to by Figure~\ref{fig:masstracks}, the mass grid is chosen to range from the hydrogen burning limit $M_H=0.08$\,$M_{\odot}$ up to $M_{\odot}$, in steps of $0.02\,M_{\odot}$ yielding 47 unique masses. In practise, we adjusted the $1\,M_{\odot}$ to $0.9999\,M_{\odot}$ since we know the Sun is inhabited.

\begin{figure}
\begin{center}
\includegraphics[width=15.5 cm]{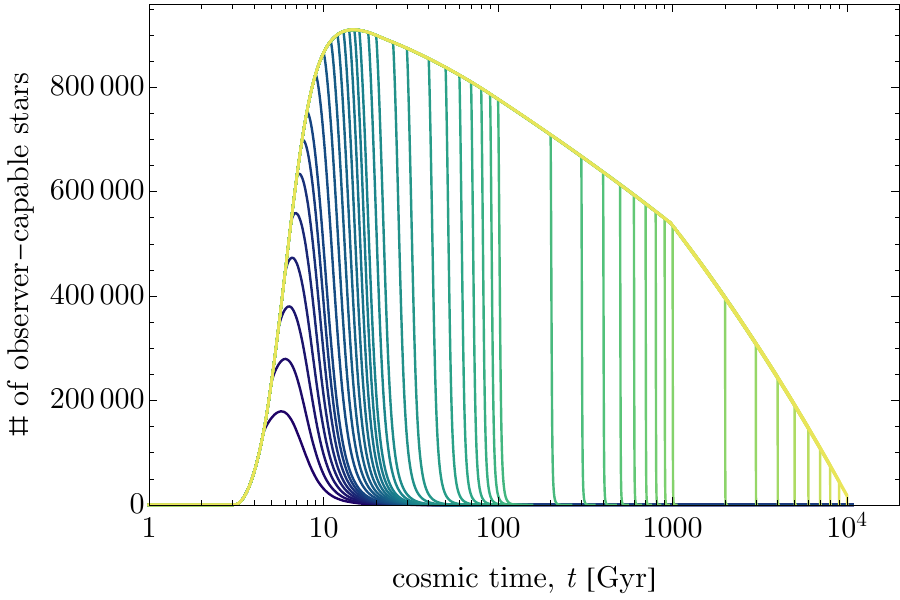}
\caption{\emph{
Number of stars capable of hosting observers (amongst a sample of $10^6$) as a function of time. The yellow track assumes that planets remain hospitable to observers during the entire main sequence lifetime of a star (effectively $\Twin=10$\,Tyr). Stepping to progressively bluer tracks, we truncate this time window $\Twin$ down to 4.5\,Gyr (bluest line).
}} 
\label{fig:twintracks}
\end{center}
\end{figure}

For $\Twin$ we adopt a quasi logarithmic scale, from $T_{\mathrm{win},\oplus}$ to $T_{\mathrm{win},\mathrm{max}}$ with 45 unique values. Analogous to Figure~\ref{fig:masstracks}, we show an example of these tracks for $\mathcal{L}_t$ holding $\Mcrit=M_H$ in Figure~\ref{fig:twintracks}.

It is perhaps surprising that the low-$\Twin$ tracks in Figure~\ref{fig:twintracks} end so early and we briefly explore that here. For example the $\Twin=10$\,Gyr track hits zero by $t\sim70$\,Gyr. Amongst our sample of $10^6$ simulated stars, the last one is born $t\sim{60}$\,Gyr. Thus, irrespective of how long the star lives (in some cases 10\,Tyr), observers are no longer possible after $(60+10)$\,Gyr because the planet dies.

Our final grid thus has 2115 unique $\{\Mcrit,\Twin\}$ combinations, for which we calculate both $\mathcal{L}_t$ and $\mathcal{L}_m$. The final grid is then interpolated with a 2D spline to evaluate these terms at arbitrary intermediate positions.

%% file: results.tex
\subsection{Joint Posterior Distribution}

\begin{figure}
\begin{center}
\includegraphics[width=15.5 cm]{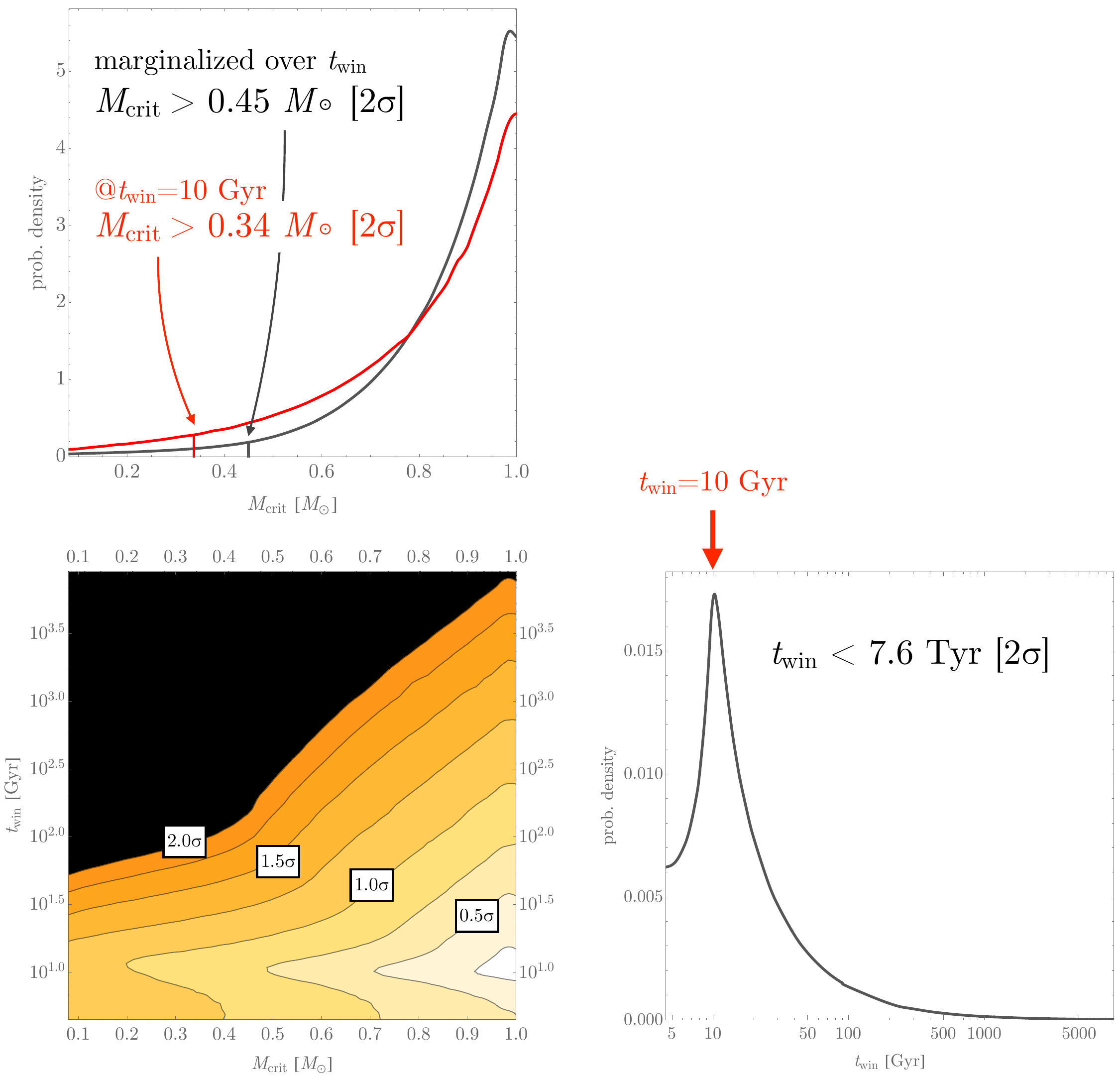}
\caption{\emph{
Corner plot of the joint posterior distribution $\pdf(\Mcrit,\Twin|t=t_0,M_{\star}=M_{\odot})$. The lower-left panel shows the joint posterior, whereas the other two panels shows the respective marginalized 1D distributions (black lines). The $\Twin$ posterior peaks at 10\,Gyr, which has good physical justification \citep{oneill:2016}. We thus show a second 1D posterior for $\Mcrit$ where $\Twin$ is held fixed at this value in red.
}} 
\label{fig:posterior}
\end{center}
\end{figure}

The joint posterior is directly given by Equation~\ref{eqn:posterior} and is plotted in the lower-left panel of Figure~\ref{fig:posterior}. We emphasize that the power of this analysis is that we allow both hypotheses to compete against one another, and indeed in collaboration with one another, to explain the data. This interaction is responsible for shaping the contours seen in Figure~\ref{fig:posterior}.However, it is often more instructive to evaluate the marginalized posteriors, found by integration:

\begin{align}
\pdf(\Twin|t=t_0,M_{\star}=M_{\odot}) &= 
\int_{\Mcrit=M_H}^{M_{\odot}} \pdf(\Mcrit,\Twin|t=t_0,M_{\star}=M_{\odot})\,\mathrm{d}\Mcrit,\\
\pdf(\Mcrit|t=t_0,M_{\star}=M_{\odot}) &= \int_{\Twin=T_{\oplus}}^{T_{\mathrm{win},\mathrm{max}}} \pdf(\Mcrit,\Twin|t=t_0,M_{\star}=M_{\odot})\,\mathrm{d}\Twin.
\end{align}

We show these in the upper-left and lower-right panels of Figure~\ref{fig:posterior}. Given the broad shape of these functions, we consider upper/lower limits derived from them to be more useful than modes/medians. Accordingly, we obtain $\Mcrit > 0.45$\,$M_{\odot}$ and $\Twin<7.6$\,Tyr to 95.45\% (2\,$\sigma$) confidence. The limit on $\Twin$ is physically unconstraining, most geological estimates far smaller values of order 5-10\,Gyr \citep{oneill:2016,cheng:2018}. In contrast, the lower limit on $\Mcrit$ excludes a stunning ${\sim}75$\% of all stars in the Universe as being potential seats for observers.

\subsection{Bayesian Model Comparison}
\label{eqn:othermodels}

In our model thus far, both hypotheses are allowed ot jointly act to explain the data. However, it is instructive to consider the case where just one of these possibilities acts - thereby revealing the explanatory power of each separately. To this end, we repeated out Bayesian inference twice, but alternating which one of the explanatory effects is disregarded. For example, if we drop the truncated window hypothesis (i.e. only the desolate M-dwarf hypothesis acts), then our posterior is simply $\pdf(\Mcrit|t=0,M_{\star}=M_{\odot})$ where we place no limit on $\Twin$ (thus one could write $\pdf(\Mcrit|t=0,M_{\star}=M_{\odot},\Twin=\infty)$). We generated new grids for this calculation since none of our original grids encompassed $\Twin=\infty$.

Similarly, we evaluated the case of dropping the desolate M-dwarf hypothesis (i.e. only the truncated window hypothesis acts) such that our posterior becomes $\pdf(\Twin|t=0,M_{\star}=M_{\odot})$ (which is equivalent to $\pdf(\Twin|t=0,M_{\star}=M_{\odot},\Mcrit=M_H)$). These tracks were already previously computed for our joint posterior and thus we simply slice the relevant cases.

Finally, we considered a fourth model, a hybrid model of the desolate M-dwarf hypothesis but with $\Twin$ fixed to $10$\,Gyr. This choice is motivated by two arguments. First, in the joint model, $\Twin$ peaks at 10\,Gyr and one can see from the covariant structure that along this slice, $\Mcrit$ is minimized. Thus, the most conservative lower limit possible on $\Mcrit$ would occur at this location. Second, 10\,Gyr falls in line with geophysical expectations argued in \citet{oneill:2016} and thus has a physically-motivated basis.

\input{evidence_tab.tex}

We next compare the upper limits on $\Mcrit$ and $\Twin$ (where applicable) from each model, and more importantly the Bayesian evidence, $\mathcal{Z}_j$, for each. Recall that the evidence is the normalization term in Equation~\ref{eqn:bayes} and represents $\pdf(\mathcal{D}|\mathcal{M}_j)$ where $\mathcal{M}_j$ is the model. Thus, the ratio of the various evidences yields the so-called Bayes factor between them. We summarize our resulting values in Table~\ref{tab:evidences}.

\subsection{Three Lessons from Model Comparison}

From this comparison, we first learn that the truncated window hypothesis, acting in isolation, is unable to well-explain the data, as it produces the worst evidence of all models considered. The Bayes factor between this hypothesis and the joint hypothesis is 32.7 (and indeed the same ratio when comparing to the desolate M-dwarf hypothesis instead). This substantially exceeds the canonical threshold of 10 defining ``strong evidence'' \citep{jeffreys:1939}. We thus conclude that some degree of mass truncation is essential to satifactorily explain the data. Of course, the reason for this is quite simple. The truncated window hypothesis has some power to explain why we emerge early in the cosmic chronology, as Figure~\ref{fig:twintracks} makes clear - but it has no ability to explain why $M_{\star}=M_{\odot}$, the Red Sky Paradox.

The second lesson revealed by our analysis is that the joint model and the desolate M-dwarf model are almost indistinguishable, with a Bayes factor of 1 to three decimal places. This again follows from the poor explanatory power of the truncated window hypothesis already discussed.

Finally, the favored model of the four is the hybrid model. This really shouldn't surprise us since to some degree it was engineered to win by virtue of fixing $\Twin$ to the modal value from the joint model. Without any other context, this would be an arbitrary choice and thus the model wouldn't even perhaps warrant discussion. However, given that \citet{oneill:2016} predict a geophysical lifetime of the Earth of this value, we argue the model is well-justified (and in any case it yields the most conservative limit on $\Mcrit$ possible). However, we point out that Bayes factor is marginal here, just 1.28 versus the joint case. Purely considering the two data points, this would not be anywhere near high enough to consider the model favored over the others. But, again, in the context of possessing a clear physical justification - we consider this the adopted modal hypothesis in what follows.

\subsection{Couldn't it all just be ``luck''?}

To complete our results section, we consider one other hypothesis, resolution I to the Red Sky Paradox \citep{redsky} - it's just luck. The easiest way to evaluate the Bayes factor of this hypothesis against our other models is to leverage the Savage-Dickey theorem \citep{dickey:1971}. Specifically, we consider here the intersection of the desolate M-dwarf hypothesis (acting in isolation) posteriors at the point where $\Mcrit \to M_H$ i.e. no mass truncation effect at all.

\begin{align}
\frac{ \mathcal{Z}_{\mathrm{luck}} }{ \mathcal{Z}_{\mathrm{desolate\,\,M}} } &= \lim_{\Mcrit \to 0} \Bigg( \frac{ \pdf(\Mcrit) }{ \pdf(\Mcrit|t=0,M_{\star}=M_{\odot}) } \Bigg).
\end{align}

Using the posterior obtained earlier in Section~\ref{eqn:othermodels}, we evaluate $\mathcal{Z}_{\mathrm{desolate\,\,M}}/\mathcal{Z}_{\mathrm{luck}} = 1597$. This is more than one order-of-magnitude in excess of the threshold for ``decisive evidence'' on the Jeffrey's scale \citep{jeffreys:1939} and thus we conclude the two puzzles cannot be trivially dismissed as happenstance.

\subsection{The Cataclysm Option}

There is another hypothesis we could consider, with some similarities to the truncated window hypothesis - a future Universe-wide cataclysm. This could come in the form of false vacuum decay \citep{devoto:2022}, the ``Grabby Aliens'' hypothesis \citep{hanson:2021} or any number of more speculative mechanisms. We consider a simplified realization of this model of the form

\begin{align}
\pdf(\tmax|t=t_0,M_{\star}=M_{\odot}) &= \frac{ \pdf(t=t_0,M_{\star}=M_{\odot}|\tmax) \pdf(\tmax) }{ \pdf(t=t_0,M_{\star}=M_{\odot}) },
\end{align}

where $\tmax$ is a simple cut-off time for observers to emerge in the Universe, where we again acknowledge the simplicity of our model is lieu of Occam's razor. Like the truncated window hypothesis, the model has no way to explain $M_{\star}=M_{\odot}$ and thus the evidence is again weak here, $\log\mathcal{Z}=-3.2563$, which is in about 2:1 disfavored over even the truncated window hypothesis. In this model, we constrain $\tmax < 310$\,Gyr (2\,$\sigma$). However, we emphasize the poor competitiveness nature of this model and do not consider is further in what follows.

%% file: evidence_tab.tex
\begin{table}
\caption{
Matrix table comparing the Bayes factors between various pairs of models, conditioned upon the hypothetical scenario that both Earth and Mars has a land fraction of $f_{\oplus}=0.292$ (and were inhabited). The column header represents the numerator and the row header represents the denominator. The upper-panel assumes a Jaynes prior, the lower-panel assumes a uniform prior in $f$.
} 
\centering 
\begin{tabular}{c c} 
\hline
Parameter & Value \\ [0.5ex] 
\hline
\textbf{Joint model} \\
\hline
$\Mcrit$ & $>0.45$\,$M_{\odot}$ [$2$\,$\sigma$]	\\
$\Twin$ & $\infty$ \\
$\log\mathcal{Z}$ & $-1.4158$ \\ [0.5ex]
\hline 
\textbf{Desolate M-dwarf hypothesis} \\
\hline
$\Mcrit$ & $>0.59$\,$M_{\odot}$ [$2$\,$\sigma$]	\\
$\Twin$ & $<7600$\,Gyr [$2$\,$\sigma$] \\
$\log\mathcal{Z}$ & $-1.4160$ \\ [0.5ex]
\hline 
\textbf{Truncated window hypothesis} \\
\hline
$\Mcrit$ & $0.08$\,$M_{\odot}$ \\
$\Twin$ & $<210$\,Gyr [$2$\,$\sigma$] \\
$\log\mathcal{Z}$ & $-2.9305$ \\ [0.5ex]
\hline 
\textbf{Hybrid hypothesis} \\
\hline
$\Mcrit$ & $>0.34$\,$M_{\odot}$	[$2$\,$\sigma$] \\
$\Twin$ & $10$\,Gyr	\\
$\log\mathcal{Z}$ & $-1.3070$ \\ [0.5ex]
\hline
\end{tabular}
\label{tab:evidences} 
\end{table}

%% file: discussion.tex
Two puzzles surface about our existence: why do we live so early in cosmic chronology and why don't we live around a typical star? We have demonstrated that the hypothesis that this is mere luck can be rejected with a Bayes factor of ${\sim}1600$. Two plausible explanations remain: M-dwarfs do not develop observers and the window for observers is truncated to some maximum timescale. We have found that the latter is untenable in isolation, rejected at a Bayes factor of ${\sim}33$, although it may contribute in collaboration with the desolate M-dwarf hypothesis.

A full joint exploration of $\Twin$ and $\Mcrit$ yields a model which is equally competitive with $\Mcrit$ acting alone. However, physical intuition guides us that planets with geological lifetimes of trillions of years (which is allowed in this model) appears in tension with geophysical expectations for the Earth \citep{oneill:2016,cheng:2018}. Our favored model is thus one where $\Twin$ is fixed to 10\,Gyr, as this not only comports with geophysical priors, but also yields the most conservative lower limit on $\Mcrit$ possible. Even so, this limit is rather stringent, excluding all stars with masses below $0.34$\,$M_{\odot}$ as developing observers to 95.45\% confidence - a cutoff which kills some two-thirds of all stars in the Universe. Dropping to $1\,\sigma$ credible interval yields $\Mcrit>0.74$\,$M_{\odot}$ and is tantamount to a Solar hegemony when it comes to observers.

We have somewhat deliberately avoided dwelling on exactly how one defines ``observer'' in this work, but strictly the implicit assumption throughout is that our experience is a representative example of ``observers''. Exactly how one interprets who is and is not representative dives into metaphysical questions beyond this author's skill to answer. Nevertheless, we argue that our work only formally applies to self-aware, reasoning observers such as ourselves, as not life in general.

Even so, if one accepts the conclusion of this work, that M-dwarfs do not develop observers, it requires some contrivance to expect simple life there still - for why shouldn't that simple life equally develop into observers eventually? What astrophysical aspect of our Sun could plausibly promote evolution over M-dwarfs? Arguably, the least contrived answer is that the stars prohibit life altogether - a question which has been ruminated upon for nearly two decades already \citep{tarter:2007,shields:2016}. The up-to-Gyr-long ZAMS lifetimes of such stars \citep{mist1,mist2}, high UV environment \citep{rugheimer:2015} and proclivity for eroding planetary atmospheres \citep{segura:2010} all offer plausible pathways for such a scenario. Indeed, recent JWST observations of TRAPPIST-1b and c are consistent with atmosphere-less worlds \citep{gillon:2025}, as well as even TRAPPIST-1e further from its star \citep{glidden:2025} (although the data remain inconclusive).

One obvious implication of our work is that M-dwarfs are not ideal targets for SETI. Given that the vast majority of stars around us are M-dwarfs \citep{reyle:2021}, excluding such objects from SETI surveys would lead to a much deeper, targeted strategy \citep{latham:1993,turnbull:2003}. Stars in the mass range of 0.74 - 1.6 Solar masses emerge as the ideal hosts from this work. Certainly an exclusive focus on such stars would be folly, intelligences beyond our own may be colonizing M-dwarfs for other purposes \citep{lingam:2023}, but again the simplest interpretation would be to de-weight them.

Finally, we reflect on the nature of Bayesian inference, such as that presented here. In comparing two models, A and B, we evaluate their ability to explain the data i.e $\pdf(\mathcal{D}|A)/\pdf(\mathcal{D}|B)$ - which is the Bayes factor. In our case, we have obtained a large Bayes factor that observers not emerging on M-dwarfs explains our data versus competing hypotheses. However, the Bayes factor is not equivalent to what might be called the odds ratio, $\pdf(A|\mathcal{D})/\pdf(B|\mathcal{D})$ - the relative probability of model A over B given the data. Such a statement necessarily requires some choice for the prior ratio $\pdf(A)/\pdf(B)$. A common approach in science is to adopt agnosticism - that all models are equally probable \textit{a-priori} \citep{trotta:2008}. But, one might argue that a Universe where all stars are inhabited is much more likely \textit{a-priori} than one inhabited solely by G-type stars, since there would be far more observers in the former case. Critically, such an extreme \textit{a-priori} preference could overwhelm the large Bayes factor. The problem with such extreme priors is that are essentially immune to data, an experimenter operating under such priors could be described as dogmatic - since their beliefs become practically impervious to observations \citep{kipping:2024}. We argue that the best way to interpret the results of work such as our own is to let go of odds factors, where anyone and everyone could inject their own personal prior beliefs. Instead, the Bayes factor is the only thing that can be objectively agreed upon\footnote{Of course, the Bayes factor is sensitive to model parameter priors still, but those can be assigned using objective reference techniques, such as the Fisher information matrix \citep{jeffreys:1939}.}. Indeed, \textit{whatever} one's model priors, the Bayes factor has a consistent interpretation: it represents the degree to which one's beliefs have changed after seeing the data. In science, this is perhaps the best we can hope for, to ask how well various hypotheses can explain the world (the data) around us. Let the philosophers argue over model priors.